\documentclass[preprint,superscriptaddress,pra,showpacs,showkeys]{revtex4}
\usepackage{graphicx,amsmath,amsfonts,amssymb}
\usepackage{times}

\begin{document}

\title{Sudden death, birth and stable entanglement in a two-qubit Heisenberg XY spin chain
\footnote{Supported by  the Natural Science Foundation of Hubei
Province, China under Grant No 2006ABA055, and the Postgraduate
Programme of Hubei Normal University under Grant No 2007D20.}}

\author{Chuan-Jia Shan\footnote{ E-mail: scj1122@163.com}}
\author{Wei-Wen Cheng}
\author{Tang-Kun Liu\footnote{Corresponding author. E-mail:
tkliuhs@163.com}}

\affiliation{College of Physics and Electronic Science, Hubei Normal
University, Huangshi 435002, China}

\author{Ji-Bing Liu}
\affiliation{College of Physics and Electronic Science, Hubei Normal
University, Huangshi 435002, China}

\affiliation{Department of Physics, Huazhong University of Science
and Technology, Wuhan 430074, China}

\author{Hua Wei}

\affiliation{State Key Laboratory of Magnetic Resonance and Atomic
and Molecular Physics, Wuhan Institute of Physics and Mathematics,
Chinese Academy of Sciences, Wuhan, 430071, China}

\date{\today}

\begin{abstract}

Taking the decoherence effect due to population relaxation into
account, we investigate the entanglement properties for two qubits
in the Heisenberg XY interaction and subject to an external magnetic
field. It is found that the phenomenon of entanglement sudden death
(ESD) as well as sudden birth(ESB) appear during the evolution
process for particular initial states. The influence of the external
magnetic field and the spin environment on ESD and ESB are addressed
in detail. It is shown that the concurrence, a measure of
entanglement, can be controlled by tuning the parameters of the spin
chain, such as the anisotropic parameter, external magnetic field,
and the coupling strength with their environment. In particular, we
find that a critical anisotropy constant exists, above which ESB
vanishes while ESD appears. It is also notable that stable
entanglement, which is independent of different initial states of
the qubits, occurs even in the presence of decoherence.
\end{abstract}



\maketitle

Entanglement, one of the essential features in quantum mechanics,
has been generally believed to be a basic resource in quantum
information.$^{[1-4]}$ In order to realize quantum-information
processors, stability of entanglement of quantum subsystems is one
of the most important premises that deserves much attention. The
entanglement has been extensively studied for various systems
including cavity-QED,$^{[5,6]}$ Ising model,$^{[7]}$ isotropic  and
anisotropic Heisenberg chains.$^{[8-11]}$ In particular, the
Heisenberg spin chain has been used to construct a quantum computer
in many physical systems such as quantum dots, nuclear spins,
electronic spins and optical lattice based systems. By proper
encoding, the Heisenberg interaction alone can support universal
quantum computation.$^{[12,13]}$ Therefore, the study of
entanglement properties of Heisenberg spin chain has received much
attention in the context of quantum information science. However, in
the real world, the environmental-induced decoherence will destroy
quantum superposition and entanglement, and thus ruin the encoded
quantum information. Counter-intuitively to conventional qubit
decoherence theory, Yu and Eberly$^{[14]}$ have shown that
entanglement may decrease abruptly to zero in a finite time due to
the influence of quantum noise,  this striking phenomenon is the
so-called entanglement sudden death (ESD). Opposite to the currently
extensively discussed ESD, entanglement sudden birth (ESB) is the
creation of entanglement where the initially unentangled qubits can
be entangled after a finite evolution  time.$^{[15]}$ Recently, many
theoretical works$^{[16-21]}$ investigated the disentanglement
dynamics in cavity-QED (Jaynes-Cummings and Tavis-Cummings model)
and spin chain. In [17], ESD induced by the effect of nonzero
initial photon number in the cavity was demonstrated via the
Tavis-Cummings model. In [18], it has been shown that atomic ESD
always occurs if ithe atomic initial state is sufficiently impure
and/or the cavity photon number is nonzero. The explicit expression
for the ESD time for various entangled states has presented in [21].
In particular, ESD has been experimentally observed recently both in
photonic qubits$^{[22]}$ and in atomic ensemble systems.$^{[23]}$

Although the environmental induced effect is not what we desired in
most cases, it has been shown that entanglement between two or more
subsystems may be induced by their collective interaction with a
common environment. A stable entangled state, in which once qubits
become entangled they will never be disentangled, was also
demonstrated in [24,25]. In this Letter, we present an exact
calculation of the  entanglement dynamics between two qubits
coupling with a common environment at zero temperature. The two
qubits interact via a Heisenberg XY interaction and are subject to
an external magnetic field. Apart from the important link to quantum
information processing, a deeper understanding of disentanglement is
also expected to provide new insights into quantum fundamentals,
particularly for quantum measurement and quantum to classical
transitions. The main purpose and motivation of the present study is
try to answer the following question: what happens to the qubits
entanglement when we consider different system parameters and
initial state in the absence or presence of the decoherence? An
important result is that  ESD and ESB appear simultaneously and is
sensitive to the initial state as well as the system parameters.
Meanwhile, a critical anisotropy constant exists, above which ESB
vanishes while ESD appears.  Moreover, it is also shown that the
decoherence due to population relaxation will always lead to stable
entanglement irrespective of the initial entangled state of the
qubits.

The Hamiltonian for an anisotropic N-qubit Heisenberg chain with
only nearest-neighbor interactions can be written as
\begin{eqnarray}
H=\sum^{N}_{i=1}(J_{x}S_{i}^{x}S_{i+1}^{x}+J_{y}S_{i}^{y}S_{i+1}^{y}+
J_{z}S_{i}^{z}S_{i+1}^{z})
\end{eqnarray}
Here we consider an anisotropic two-qubit Heisenberg XY system
coupled to an environment in an external magnetic field $\omega$
along the z-axis, the corresponding Hamiltonian reads
\begin{eqnarray}
H=J(S_{1}^{+}S_{2}^{-}+S_{1}^{-}S_{2}^{+})+\Delta
(S_{1}^{+}S_{2}^{+}+S_{1}^{-}S_{2}^{-})+\omega (S_{1}^{z}+S_{2}^{z})
\end{eqnarray}
where $J=(J_{x}+J_{y})/2$, $\Delta=(J_{x}-J_{y})/2$, and $S^{\pm
}=S^{x}\pm i S^{y}$ are the spin raising and lowering operators, the
parameter $\Delta$ describes the spatial anisotropy of the spin-spin
interaction. The anisotropy parameter can be controlled by varying
$J_{x}$ and $J_{y}$, which may possibly be achieved for an optical
lattice system$^{[26]}$, the effective external magnetic field is
defined by the energy levels of our qubits. The description of the
time evolution of an open system is provided by the master equation,
which can be written most generally in the Lindblad form with the
assumption of weak system-reservoir coupling and Born-Markov
approximation. The Lindblad equation for our case thus reads
\begin{eqnarray}
\frac{d\rho }{dt}=-i[H,\rho ]+\gamma
\sum\limits_{j=1,2}[S_{j}^{-}\rho S_{j}^{+}-\frac{1}{2}\left\{
S_{j}^{+}S_{j}^{-},\rho \right\}]
\end{eqnarray}
where $\gamma$ is the relaxation rate of the qubits and we have
assumed them to be the same; the assumption is reasonable provided
the interaction does not significantly alter the energy level
separations. $\{\}$ means anticommutator.

Firstly, the solution of Eq. (3) depends on the initial state of the
qubits, and we assume that the initial state of the system is in a
general form
$\cos\theta|\downarrow\downarrow\rangle+\sin\theta|\uparrow\uparrow\rangle$.
We note that, for the class of the initial states considered here,
the solution of Eq.(3) has the matrix form
\begin{eqnarray} \rho=\left(
\begin{array}{cccc}
\rho _{1,1} & 0 & 0 & \rho _{1,4} \\
0 & \rho _{2,2} & \rho _{2,3} & 0 \\
0 & \rho _{3,2} & \rho _{3,3} & 0 \\
\rho _{4,1} & 0 & 0 & \rho _{4,4}%
\end{array}\right)
\end{eqnarray}
in the  two-qubit product state basis of $\{\left\vert
\uparrow \uparrow \right\rangle ,\left\vert \uparrow \downarrow
\right\rangle ,\left\vert \downarrow \uparrow \right\rangle
,\left\vert \downarrow \downarrow \right\rangle \}$.

Since decoherence process leads the pure quantum system state to
mixed states, we use the concurrence as a measure of entanglement.
The concurrence $C = 0$ corresponds to a separable state and $C = 1$
to a maximally entangled state. Nonzero concurrence means that the
two qubits are entangled. Using Wootters¡¯ formula,$^{[26]}$ for a
system described by the above density matrix in Eq. (4), the
concurrence is
\begin{eqnarray}
C(\rho )=\max (0,\lambda _{1}-\lambda _{2}-\lambda _{3}-\lambda
_{4})
\end{eqnarray}
where $\lambda _{1}$, $\lambda _{2}$, $\lambda _{3}$, $\lambda _{4}$
are the eigenvalues in a decreasing order of the spin-flipped
density operator $R$ defined by $R=\sqrt{\sqrt{\rho }\tilde{\rho}
\sqrt{\rho }}$ with $\tilde{\rho} =(\sigma _{y}\otimes \sigma
_{y})\rho ^{\ast }(\sigma _{y}\otimes \sigma _{y})$, $\tilde{\rho}$
denotes the complex conjugate of $\rho$, $\sigma _{y}$ is  the usual
Pauli matrix. Then the concurrence can be expressed as
\begin{eqnarray}
C=max[0, 2(\sqrt{\rho_{23}\rho_{23}}-\sqrt{\rho_{11}\rho_{44}}),
2(\sqrt{\rho_{14}\rho_{41}}-\sqrt{\rho_{22}\rho_{33}})]
\end{eqnarray}
In the following, we use this formalism to investigate the
entanglement dynamics and decoherence  under different system
parameters, such as the anisotropic parameter, external magnetic
field, for several different initial cases: the disentangled of the
two qubits ($\theta=0,\theta=\frac{\pi}{2}$), not maximal entangled
state ($\theta=\frac{\pi}{8}$) and maximal entangled state
($\theta=\frac{\pi}{4},\theta=\frac{3\pi}{4}$).

In Figs.1 and 2, the time evolution of the concurrence is plotted
for various values of the external magnetic field parameter $\Omega$
with and without the decoherence when the qubits are initially in
the different initial state. From Figs.1(a) and 1(c), one can find
that the entanglement evolves periodically in the absence of the
decoherence. If there is no magnetic field, we can see that
unentangled initial state periodically generates maximally entangled
states; while the maximally entangled initial state does not evolve
in time as shown in Fig.1(c), as it is an eigenvector of the
Hamiltonian in the absence of the surrounding environment. The solid
line represents result when the control field is turned off,  the
dashed, dotted and dash-dotted lines correspond to different control
field strength $\omega=0.2, 0.6,$ and $1.0$, respectively. In
contrast to the solid line in Fig.1(a), once the magnetic field is
given, the amplitude of these oscillations decreases with the
increase of the external magnetic field. It makes slow oscillation
around  the maximal value of concurrence, $C=1$. Similar behaviours
to those in Figs.1(a) and 2(c) are shown in Figs.2(a) and 2(c). From
Fig.1(b), we can see that the two-qubit state can evolve into a
stationary entangled state under the collective decay from initial
unentangled state. In other words, decoherence drives the qubits
into a stationary entangled state instead of completely destroying
the entanglement. Moreover, the stationary entanglement of two
qubits increases with the decrease of the external magnetic field.
Therefore, this can provide us a feasible way to manipulate and
control the entanglement by changing the external magnetic field.
Contrarily, Figures 1(d) and 2(d) show that as the external magnetic
field increases the entanglement of two qubits can fall abruptly to
zero, and will recover after a period of time. Therefore, the ESD
appears and is related to both the initial state and the external
magnetic field. Even though the initial system has the same
entanglement, different evolution will appear.  When the two spins
are initially prepared in their excited state, i.e.
$\theta=\frac{\pi}{2}$ , the result is quite different from that of
Fig.1(b). The concurrence versus parameter $t$ is plotted in Fig.
2(b), indicating a threshold value of parameter $t$, only above
which concurrence begins to be nonzero, i.e. the quantum correlation
starts to appear. This is the so-called ESB and the delayed time for
the appearance of ESB increases with the increasing of the strength
of external magnetic field.

At this stage, we turn to study the influence of anisotropy effect
of the system on the entanglement dynamics. Figure 3 illustrates the
time evolution of the concurrence for different values of the
anisotropy constant $\Delta$ with decoherence effect. By comparing
Fig.3(c) and 3(d), it is observed that at a fixed external magnetic
field,  a critical anisotropy constant exists, above which ESB
vanishes while ESD appears. It is the anisotropy of interaction that
leads to considerable difference in entanglement evolution, hence
entanglement is rather sensitive to any small change of the system
anisotropy. That is to say by adjusting the anisotropic constant
alone one can also obtain both ESD and ESB. Another important
property revealed by Fig.3(c) is that ESD occurs twice with the
increase of the system anisotropy. In the case of weak external
magnetic field and strong anisotropy interaction, as we show in the
dash-dotted lines of Figs.3(a), 3(b) and 3(c), concurrence actually
goes abruptly to zero in a finite time and remains zero thereafter,
i.e., the ESD will always survive in a strong anisotropic
interaction. At strong control field in Fig.3(d), the system loses
its entanglement completely for a short period of time, and then it
is entangled again some time later. Thus the lifetime of ESD can be
controlled in our model by applying local external magnetic field.
Despite the presence of decoherence, the results in Fig. 4 show that
the concurrence reaches the same steady value, after some
oscillatory behaviour, for a given set of system parameters
regardless of the initial state of the system. At $T = 0$, the
corresponding steady concurrence is found to be
\begin{eqnarray}
C_{s}=\frac{2\Delta\sqrt{4\omega^{2}+\Delta^{2}}-2\Delta^{2}}{4(\omega^{2}+\Delta^{2})+\gamma^{2}}
\end{eqnarray}
The steady-state concurrence is seen to depend on the system
parameters $\omega, \Delta$, and $\gamma$ while independent of the
$J$ and the initial entangled state of the qubits.

In summary,  we have presented an analytic solution for the
evolution of entanglement for different initial system states. It is
found that ESD and ESB appear simultaneously, depending on the
initial state, the anisotropic parameter, external magnetic field,
and the coupling strength with the environment. A stable
entanglement, controllable by the values of the system parameters,
will always be obtained for zero or finite. Our results will shed
light on understanding of entanglement dynamics of quantum systems
with environmental effect as well as  the ESD and ESB in a
correlated
environment.\\
\indent We thank Z-Y Xue for his reading of the manuscript. \\
\indent  \emph{Note added}- After the completion of this paper, C.
E. L\'{o}pez brought to our attention the Letter published in [28]
and we thank C. E. L\'{o}pez for useful discussions.

\newpage
\begin{figure}
\begin{center}
\includegraphics[width=1.0\textwidth]{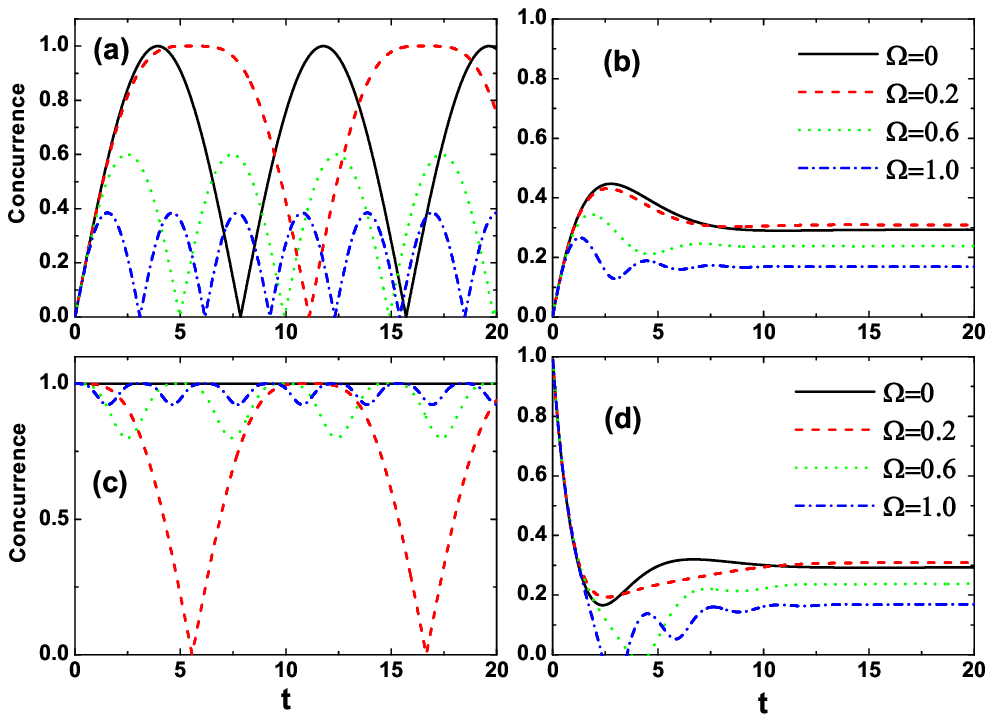}\\
\caption{The time evolution of the concurrence for various values of
the external magnetic field parameter $\Omega$ with $\Delta=0.2$
when the qubits are initially in different initial state.
(a)$\theta=0, \gamma=0$, (b)$\theta=0, \gamma=0.5$,
(c)$\theta=\frac{\pi}{4}, \gamma=0$, (d)$\theta=\frac{\pi}{4},
\gamma=0.5$.}\label{Fig.1.EPS}
\end{center}
\end{figure}

\begin{figure}
\begin{center}
\includegraphics[width=1.0\textwidth]{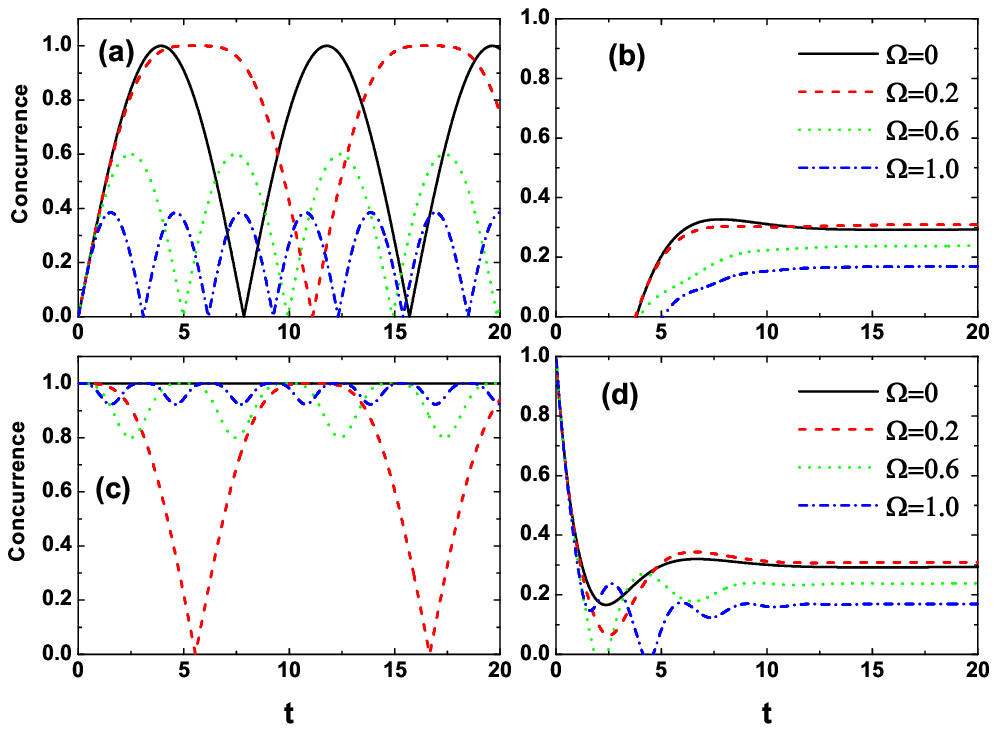}\\
\caption{The time evolution of the concurrence for various values of
the external magnetic field parameter $\Omega$ with $\Delta=0.2$
when the qubits are initially in different initial state.
(a)$\theta=\frac{\pi}{2}, \gamma=0$, (b)$\theta=\frac{\pi}{2},
\gamma=0.5$, (c)$\theta=\frac{3\pi}{4}, \gamma=0$,
(d)$\theta=\frac{3\pi}{4}, \gamma=0.5$.}\label{Fig.2.EPS}
\end{center}
\end{figure}

\begin{figure}
\begin{center}
\includegraphics[width=1.0\textwidth]{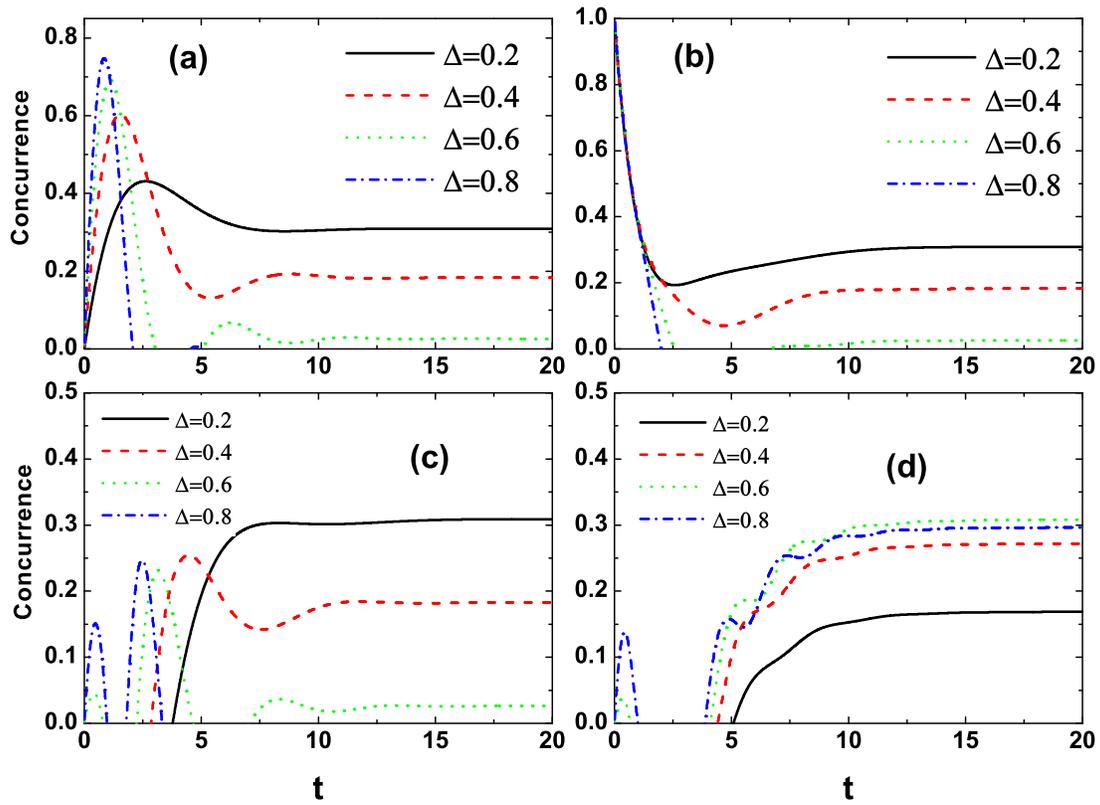}\\
\caption{The time evolution of the concurrence for various values of
the anisotropic parameter with $ \gamma=0.5$ when the qubits are
initially in different initial state. (a)$\theta=0, \Omega=0.2$,
(b)$\theta=\frac{\pi}{4}, \Omega=0.2$, (c)$\theta=\frac{\pi}{2},
\Omega=0.2$, (d)$\theta=\frac{\pi}{2}, \Omega=1.0$}\label{Fig.3.EPS}
\end{center}
\end{figure}

\begin{figure}
\begin{center}
\includegraphics[width=0.6\textwidth]{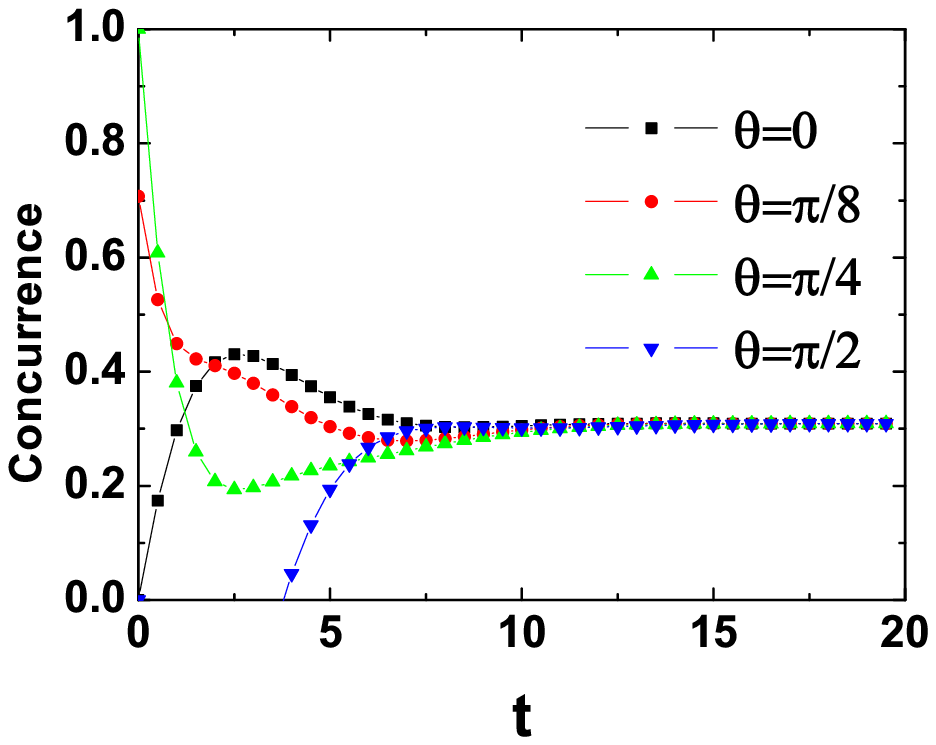}\\
\caption{The time evolution  of the concurrence with $\Omega=0.2,
\Delta=0.2,\gamma=0.5$ when the qubits are initially in different
initial state. }\label{Fig.4.EPS}
\end{center}
\end{figure}


\begin{thebibliography}{100}
\bibitem{1} Bennett C H et al 1993 \emph{Phys. Rev. Lett.} \textbf{70} 1895\\
            Xue Z Y, Yang M, Yi Y M, and Cao Z L 2006 \emph{Opt. Commun.} \textbf{258} 315

\bibitem{2} Wei H, Deng Z J, Zhang X L and Feng M 2007 \emph{Phys. Rev. A} \textbf{76}
054304\\ Wei H, Fang R R, Liu J B et al 2008 \emph{J. Phys.
B}\textbf{41} 085506

\bibitem{3} Xue Z Y, Yi Y M, and Cao Z L 2007 \emph{Physica A} \textbf{374}
119\\
           Xue Z Y, Yi Y M, and Cao Z L 2006 \emph{J. Mod. Opt.} \textbf{53}
2725

\bibitem{4} Grover L 1998 \emph{Phys. Rev. Lett.} \textbf{80} 4329

\bibitem{5} Zheng S B and Guo G C 2000 \emph{Phys. Rev. Lett.} \textbf{85} 2392

\bibitem{6} Liu T K, Cheng W W, Shan C J et al  2007 \emph{Chin. Phys.} \textbf{16} 3697

\bibitem{7} Pang C Y and Li Y L 2006 \emph{Chin. Phys. Lett.} \textbf{23}
3145

\bibitem{8} Wang X G 2001 \emph{Phys. Rev. A} \textbf{64} 012313; 2002 \textbf{66} 044305; 2002 \textbf{66} 034302

\bibitem{9} Zhang G F and Li S S 2005 \emph{Phys. Rev. A}  \textbf{72} 034302

\bibitem{10}Zhou L, Song H  S, Guo Y Q and  Li C 2003 \emph{Phys. Rev. A}
\textbf{68}024301

\bibitem{11} Shan C J, Cheng W W, Liu T K, Huang Y X and Li H 2008  \emph{Chin. Phys. Lett.} \textbf{25} 817;
\emph{Chin. Phys.} \textbf{17} 0794; Cheng W W, Huang Y X, Liu T K
and Li H 2007 \emph{Physica E} \textbf{39} 150

\bibitem{12} Christandl M,  Datta N, Ekert A and Landahl A J 2004 \emph{Phys. Rev. Lett.} \textbf{92}
187902

\bibitem{13} Mohseni M and Lidar D A 2005 \emph{Phys. Rev. Lett.} \textbf{94} 040507

\bibitem{14} Yu T and Eberly J H 2004 \emph{Phys. Rev. Lett.} \textbf{93} 140404

\bibitem{15} Ficek Z and Tanas R 2008 \emph{Phys. Rev. A} \textbf{77} 054301

\bibitem{16} Yu T and Eberly J H 2006 \emph{Phys. Rev. Lett.} \textbf{97} 140403

\bibitem{17} Shan C J, Xia Y J 2006 \emph{Acta. Phys. Sin.} \textbf{55} 1585

\bibitem{18} Man Z X,  Xia Y J and An N B 2008 \emph{J. Phys. B} \textbf{41} 085503

\bibitem{19} Yang Q, Yang M, Cao Z L et al 2008 \emph{Chin. Phys. Lett.} \textbf{25} 825

\bibitem{20} Jing J,  L\"{u} Z G and  Yang G H 2007 \emph{Phys. Rev. A} \textbf{76}
032322
\bibitem{21} Ikram M, Li Fl and Zubairy M S 2007 \emph{Phys. Rev. A} \textbf{75} 062336

\bibitem{22} Almeida M P et al 2007 \emph{Sience} \textbf{316} 579

\bibitem{23} Laurat J et al 2007 \emph{Phys. Rev. Lett.} \textbf{99} 180504

\bibitem{24} Hartmann L, D\"{u}r W and  Briegel H J 2007 \emph{Phys. Rev. A}
\textbf{74} 052304

\bibitem{25} Abliz A et al 2006 \emph{Phys. Rev. A} \textbf{74} 052105

\bibitem{26}S{\o}rensen A and M{\o}lmer K 1999 \emph{Phys. Rev. Lett.} \textbf{83}
2274; Duan L M, Demler E, Lunkin M D 2003 \emph{Phys. Rev. Lett.}
\textbf{91} 090402

\bibitem{27}Wooters W K 1998 \emph{Phys. Rev. Lett.} \textbf{80} 2245

\bibitem{28}L\'{o}pez C E, Romero G, Lastra F, Solano E, and  Retamal J C 2008 \emph{Phys. Rev. Lett.} \textbf{101} 080503
\end{thebibliography}
\end{document}